# Single Plasmon Switching with *n* Quantum Dots System Coupled to One-Dimentional Waveguide

Nam-Chol Kim, [†,*] Myong-Chol Ko,[†] Song-Jin Im, [†] Qu-Quan Wang[‡]

[†] Department of Physics, Kim Il Sung University, Pyongyang, D. P. R. Korea

[‡] Department of Physics, Key Laboratory of Acoustic and Photonic Materials and Devices of Ministry of Education, Wuhan University, Wuhan 430072, P. R. China

**Abstract:** Switching of a single plasmon interacting with *n* equally spaced quantum dots coupled to one-dimensional surface plasmonic waveguide is investigated theoretically via the real-space approach. We showed that the transmission and reflection of a single plasmon can be switched on or off by dynamically tuning and changing the number of the equal transition frequencies of the QDs. Increasing the number of doped QDs having the equal transition frequencies results in the wide – band reflection of a single plasmon and the complete transmission peaks near the resonant frequencies.

[*] Electronic mail: elib.rns@hotmail.com

The interaction between light and matter has always been a fundamental topic in physics, and its most elementary level is the interaction of a single photon to a single emitter.[1] Since photons are ideal carriers of quantum information, photons are naturally considered to replace electrons in future information technology.[2] Controlling single-photon transport has attracted particular interests for some fundamental investigations of photon-atom interaction and for its applications in quantum information. Recently, theoretical idea of a single-photon transistor has also emerged.[3] Many theoretical[4–11] and experimental[12–15] works reported the photon emission statistics and the photon scattering in different quantum systems. Most proposals for a single photon transport are based on the real-space method[4,5] In the previous studies, the authors have mainly considered the scattering properties of a single photon interacting with an emitter, and they reported that a strong modification of photon transmission spectra could be achieved. Recently, the interaction between a single photon and several emitters are also investigated[16–18] and the authors have shown the peculiar scattering properties of a single photon, quite different from the case of one emitter. However, they mainly focus on the case where the quantum emitters are all the same.

In this paper, we investigate theoretically the single propagating plasmon scattered by equally spaced $n$ quantum dots (QDs) coupled to a one-dimensional (1D) surface plasmatic waveguide, where the transition frequencies of the QDs can be different with each other.

The coupling between metal nanowires and emitters is important for tailoring light-matter interactions, and for various potential applications.[19] The nanowire exhibits good confinement and guiding even when its radius is reduced well below the optical wavelength. In this limit, the effective Purcell factor $P \equiv \Gamma_{pl} / \Gamma'$ can exceed $10^3$ in realistic systems according to the theoretical results,[20] where $\Gamma_{pl}$ is the spontaneous emission rate into the surface plasmons(photons) and $\Gamma'$ describes contributions from both emission into free space and non-radiative emission via ohmic losses in the conductor. Both experimental and theoretical investigations show that the emission properties of QDs can be significantly modified near the metallic nanostructures.[21, 22] Recent investigations[23] have further extended into the regime of interactions of quantum emitters such as QDs and propagating SPs. Motivated by these considerations, we investigate the scattering of a single plasmon interacting with *n* QDs coupled to 1D surface plasmonic waveguide which is a metal nanowire [Fig. 1].

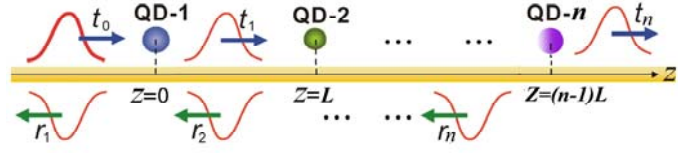

Fig.1 (Color online). Schematic diagram of a system consisting of a single plasmon and *n* equally spaced QDs coupled to a metal nanowire. $t_i$ and $r_i$ are the transmission and reflection amplitudes at the place $z_i$, respectively.

Under the rotating wave approximation, the Hamiltonian of the system in real space is given by[4]

$$H = \sum_{j=1}^{n}\left[(\omega_e^{(j)} - i\Gamma'_j/2)\sigma_{ee}^j + \omega_g^{(j)}\sigma_{gg}^j\right] + \sum_k \omega_k a_k^+ a_k + \sum_{j=1}^{n}\sum_k g_j\left(a_k^+ \sigma_{ge}^j + a_k \sigma_{eg}^j\right) \quad (1)$$

where $\omega_e^{(j)}$ and $\omega_g^{(j)}$ are the eigenfrequencies of the state $|g\rangle$ and $|e\rangle$ of the *j* th QD, respectively, $\omega_k$ is the frequency of the surface plasmon with wavevector $k$ ($\omega_k = v_g|k|$). $\sigma_{ge}^j = |g\rangle_{jj}\langle e|$ ($\sigma_{eg}^j = |e\rangle_{jj}\langle g|$) is the lowing (raising) operators of the *j*th QD, $a_r^+(z_j)$ ($a_l^+(z_j)$) is the bosonic operator creating a right-going (left-going) plasmon at position $z_j$ of the *j*th QD. $v_g$ is the group velocity of the surface plasmon, the non-Hermitian term in *H* describes the decay of state $\omega_e^{(j)}$ at a rate $\Gamma'_j$ into all other possible channels. $g_j = (2\pi\hbar/\omega_k)^{1/2}\Omega_j \mathbf{D}_j \cdot \mathbf{e}_k$ is the coupling constant of the *j*th QD with surface plasmon, $\Omega_j$ is the resonance energy of the *j*th QD, $\mathbf{D}_j$ is the dipole moment of the *j*th QD, $\mathbf{e}_k$ is the polarization unit vector of the surface plasmon[4].

Assuming that a plasmon is incoming from the left with energy $E_k = \hbar\omega_k$, then the eigenstate of the system, defined by $H|\psi_k\rangle = E_k|\psi_k\rangle$, can be constructed in the form

$$|\psi_k\rangle = \int dz\left[\phi_{k,r}^+(z)a_r^+(z) + \phi_{k,l}^+(z)a_l^+(z)\right]|0,g\rangle + \sum_{j=1}^{n} e_k^{(j)}|0,e_j\rangle, \quad (2)$$

where $|0, g\rangle$ denotes the vacuum state with zero plasmon and *n* QDs being unexcited, $|0, e_j\rangle$ denotes the vacuum field and only the *j*the QD in the excited state and $e_k^{(j)}$ is the probability amplitude of the *j*th QD in the excited state. $\Phi_{k,r}^+(z)$ ($\Phi_{k,l}^+(z)$) is the wavefunction of a right-going (a left-going) plasmon at position *z*.

For a plasmon incident from the left, the mode functions $\Phi^+_{k,r}(z)$ and $\Phi^+_{k,l}(z)$ take the forms $\phi^+_{k,r}(z<0)=e^{ikz}$, $\phi^+_{k,r}((j-1)L<z<jL)=t_j e^{ik(z-jL)}$, $\phi^+_{k,r}(z>(n-1)L)=t_n e^{ik(z-nL)}$, $\phi^+_{k,l}(z<0)=r_1 e^{-ikz}$, $\phi^+_{k,l}((j-1)L<z<jL)=r_{j+1}e^{-ik(z-jL)}$, and $\phi^+_{k,l}(z>(n-1)L)=0$, respectively, where $j = 1, 2, \cdots, n$ and $L$ is the spacing between the neighboring QDs. Here $t_j$ and $r_j$ are the transmission and reflection amplitudes at the place $z_j$, respectively. By substituting Eq. (2) into $H|\psi_k\rangle = E_k|\psi_k\rangle$, we obtain a set of equations as:

$iv_g[\partial_z \phi^+_{k,l}(z_j)] + g_j e^{(j)}_k = \omega_k \phi^+_{k,l}(z_j)$, $-iv_g[\partial_z \phi^+_{k,r}(z_j)] + g_j e^{(j)}_k = \omega_k \phi^+_{k,r}(z_j)$, $(\Omega_j - i\Gamma'_j/2)e^{(j)}_k + g_j(\phi^+_{k,r}(z_j) + \phi^+_{k,l}(z_j)) = \omega_k e^{(j)}_k$, where $\omega^{(j)}_g = 0$, $\omega^{(j)}_e - \omega^{(j)}_g = \Omega_j$ ($j=1, 2, \cdots, n$).

By taking the boundary conditions of the mode functions $t_{j-1} + r_j = t_j e^{-ikL} + r_{j+1} e^{ikL}$, where $t_0 = 1$, $r_{n+1} = 0$ and $j = 1, 2, \cdots, n$, into account, we obtain the transmission and the reflection amplitudes, respectively, as ($J_j = g^2_j/v_g$, $\Delta^{(j)}_k = \Omega_j - \omega_k$ ($j = 1, 2, \cdots, n$))

$t_j e^{-ikL} + (iJ_j/\Delta_j - 1)t_{j-1} + iJ_j r_j/\Delta_j = 0$, $iJ_j t_{j-1}/\Delta_j + (iJ_j/\Delta_j + 1)r_j - r_{j+1}e^{ikL} = 0$,

where $\Delta \equiv i\Gamma'_j/2 - \Delta^{(j)}_k$.

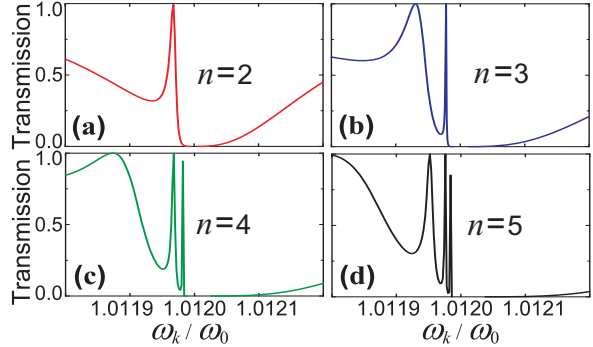

Fig. 2 (Color online). Transmission spectra of a propagating plasmon interacting with $n$ QDs ($\Omega_i = 1.0125\omega_0$, $i = 1, 2, \cdots, n$) versus $\omega_k$ for phases $kL = 0.1\pi$, where the frequency is in units of $\omega_0 \equiv 2\pi v_g/L$ and $J = 10^{-4}\omega_0$: (a) $n = 2$, (b) $n = 3$, (c) $n = 4$, (d) $n = 5$.

The scattering property of the single plasmon in the long time limit is characterized by the transmission coefficient $T_j \equiv |t_j|^2$ and reflection coefficient $R_1 \equiv |r_1|^2$. Figure 2 shows typical transmission spectra for phase $kL = 0.1\pi$. In all our calculaltions, we suppose that $\Gamma'_i = 0$ and $J_i = J = $ const ($j = 1, 2, \cdots, n$). In the case of a single emitter problem, the transmission is very small near the resonance and is exactly zero at the resonance,[4] which is quite different from the cases $n > 1$ as shown in Fig. 2. Calculation shows that there are $n - 1$ transmission peaks near the resonance frequency, which is shown in Fig. 2 for the cases with few QDs. The position of the complete transmission peak can be determined from the conditions as $r_1 = 0$.

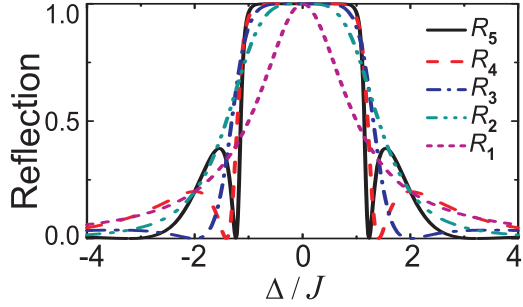

Fig. 3 (Color online). Reflection spectra of a propagating plasmon interacting with $n$ equally spaced QDs as a function of $\Delta / J$ for $kL = 0.5\pi$, where $\Delta_i = \Delta$ ($i = 1, 2, \cdots, n$), $\omega_0 \equiv 2\pi v_g / L$ and $J = 10^{-4}\omega_0$.

Figure 3 shows the reflection spectra of a propagating plasmon interacting with $n$ equally spaced QDs as a function of $\Delta / J$ for $kL = 0.5\pi$, where all the QDs have the same transition frequencies. There appears only one complete reflection peak and there is no transmission peak when $n = 1, 2$, which is quite different from the case that $n > 2$. From the Fig. 3, we can see easily that there appear complete transmission peak near the zero detuning, $|\Delta| = 2, 1.42, 1.24$, when $n = 3, 4, 5$, respectively. As we can see easily from the Fig. 3, the width of complete reflection peak of the single photon scattering becomes wider, as the number of doped QDs is increased, which is quite interesting result. In the case of a single emitter[4, 5], there appears a complete reflection only in a specific frequency, i. e., resonant frequency, which results in the physical difficulties in practical applications, because the optical pulse controlled is actually a superposition of the plan wave with different frequencies where the off-resonant components could deviate from the complete reflection dramatically. Our result shows that increasing the number of QDs results in broadening the band of complete reflection. Especially, when $n = 5$, the complete reflection peak is nearly rectangular form, and the region of wide-band of complete reflection reachese near the saturable maximum, which could find practical applications.

Now, we consider the scattering spectrum of a single plasmon coherently interacting with several QDs, the transition frequencies of which could be different with each other. Firstly, we investigate the case $n = 3$, the two transition frequencies of which are equal to each other as shown in Fig. 4. Figures 4(a) and 4(b) show transmission spectrum when the second and third transition frequencies are equal to each other, $\Omega_2=\Omega_3$, from which the complete transmission peak appears near the frequency resonant with that of the two same QDs. This result stands also for the case $\Omega_1=\Omega_3$, but the width of the transmission peak is not so sharp as that of the case $\Omega_2=\Omega_3$ [Fig. 4(c) and 4(d)]. The inversion of the order of QDs doesn't change the scattering spectrum of a single propagating plasmon at all, as shown in Figs. 4(a)-4(b) and 4(e)-4(f).

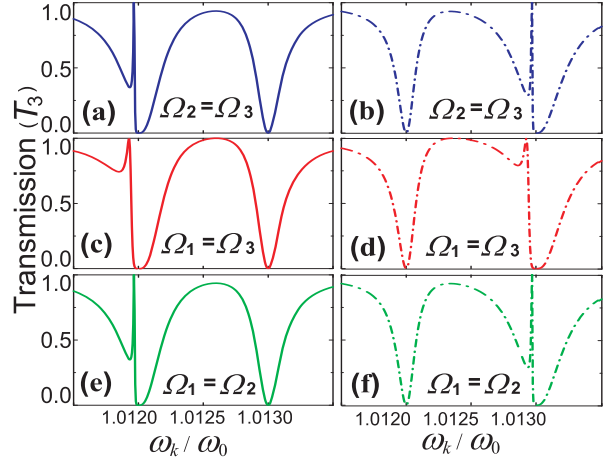

Fig. 4 (Color online). Transmission spectra of a propagating plasmon interacting with three QDs versus $\omega_k$ for $kL = 0.1\pi$, where the transition frequencies of the two QDs are equal to each other: (a) $\Omega_1=1.0130\omega_0$, $\Omega_2=\Omega_3=1.0120\omega_0$, (b) $\Omega_2=1.0120\omega_0$, $\Omega_1=\Omega_3=1.0130\omega_0$, (c) $\Omega_2=1.0130\omega_0$, $\Omega_1=\Omega_3=1.0120\omega_0$, (d) $\Omega_2=1.0120\omega_0$, $\Omega_1=\Omega_3= 1.0130\omega_0$, (e) $\Omega_3=1.0130\omega_0$, $\Omega_1= \Omega_2=1.0120\omega_0$, (f) $\Omega_3=1.0120\omega_0$, $\Omega_1= \Omega_2=1.0130\omega_0$, where $\omega_0 \equiv 2\pi v_g / L$ and $J = 10^{-4}\omega_0$.

Next, we consider the case of $n = 4$, where the two pairs of transition frequencies are equal to $1.0120\omega_0$ and $1.0130\omega_0$, respectively, as shown in Fig. 5. The order of QDs influences the scattering spectrum, as can be seen in Fig. 5. The scattering spectra of such cases as $\Omega_1 = \Omega_2$ ($\Omega_3 = \Omega_4$), $\Omega_1 = \Omega_3$ ($\Omega_2 = \Omega_4$), $\Omega_1 = \Omega_4$ ($\Omega_2 = \Omega_3$) are different with each other, among which the complete transmission peak in the first case ($\Omega_1 = \Omega_2$, $\Omega_3 = \Omega_4$) is most sharp and therefore is very useful for a single plasmon switching. Anyway, the scattering spectrum considered above is quite different with those of only two QDs system, where there is no complete transmission peak near the resonant frequencies[17]. Our calculation shows the existence of the equal transition frequencies results in the complete transmission near the resonant frequencies, which is valuable for the practical applications.

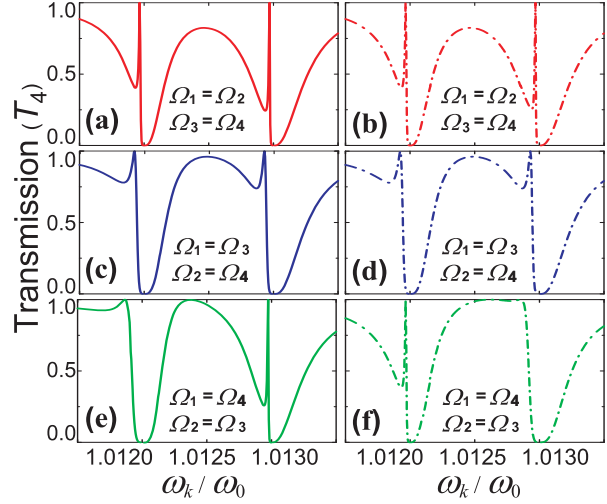

Fig. 5 (Color online). Transmission coefficients of a propagating plasmon interacting with four QDs versus $\omega_k$ for $kL = 0.1\pi$: (a) $\Omega_1=\Omega_2=1.0120\omega_0$, $\Omega_3=\Omega_4=1.0130\omega_0$, (b) $\Omega_1=\Omega_2=1.0130\omega_0$, $\Omega_3=\Omega_4=1.0120\omega_0$, (c) $\Omega_1=\Omega_3=1.0120\omega_0$, $\Omega_2=\Omega_4=1.0130\omega_0$, (d) $\Omega_1=\Omega_3=1.0130\omega_0$, $\Omega_2=\Omega_4=1.0120\omega_0$, (e) $\Omega_1=\Omega_4=1.0120\omega_0$, $\Omega_2=\Omega_3=1.0130\omega_0$, (f) $\Omega_1=\Omega_4=1.0130\omega_0$, $\Omega_2=\Omega_4=1.0120\omega_0$, where $\omega_0 \equiv 2\pi v_g / L$ and $J = 10^{-4}\omega_0$.

It is also interesting to consider the cases that all the transition frequencies except one are equal to each other. When there are two frequencies of $n$ QDs, there appear two complete reflection peaks, one of which is very narrow and the another is wider. The narrow complete reflection peak is corresponding to a single transition frequency and the wide complete reflection peak is corresponding to the other transition frequency of $n$ -1 QDs. In Fig. 6, the solid curve (blue), the dash-doted curve (red) and the dash-dot-doted curve (green) represent the scattering spectrum of the cases $n = 5$ ($\Omega_5=1.0130\omega_0$, the others equal to $1.0120\omega_0$), 4($\Omega_4=1.0130\omega_0$, the others equal to $1.0120\omega_0$), 3($\Omega_3=1.0130\omega_0$, the others equal to $1.0120\omega_0$) respectively. For example, the solid curve represents the reflection coefficient of the scattering of a single plasmon interacting with $n = 5$ QDs, where $\Omega_1 = \Omega_2 = \Omega_3 = \Omega_4=1.0120\omega_0$, $\Omega_5=1.0130\omega_0$. As can be seen in Fig. 6, the complete reflection region near the frequency $1.0130\omega_0$ is only in a specific point, while the region near the frequency $1.0120\omega_0$ with which the $n$ -1 QDs exist is extended to a wide band. Furthermore, as the number of QDs having the equal transition frequency results in the changing of the width of reflection peaks, which suggests the possibility of controlling the width of the band for the complete reflection region. Anyway, our calculation shows that increasing the number of doped QDs in the system results in the broadening of the band of reflection spectrum, results in the controllable switching of a single plasmon not only in a specific frequency, but also in wide-band frequency region, the latter case is very valuable for the practical applications.

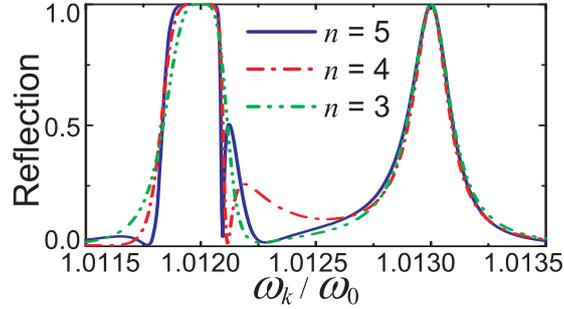

Fig. 6 (Color online). Reflection spectra of a propagating plasmon interacting with $n$ QDs versus $\omega_k$ for $kL = 0.6\pi$, where all the transition frequencies except one are equal to each other and $n = 3$ (green dash-dot-doted curve), 4 (red dash-doted curve), and 5 (blue dashed curve), respectively. In all cases, we set $\omega_0 \equiv 2\pi v_g / L$ and $J = 10^{-4}\omega_0$.

In summary, we investigated theoretically the scattering properties of a single plasmon interacting with $n$ equally spaced QDs which are coupled one dimensional plasmatic waveguide. We showed that the transmission and reflection of a single plasmon can be switched on or off by dynamically tuning and changing the number of the equal transition frequencies of the QDs. We also note that increasing the number of the equal transition frequencies of the QDs results in the wide – band reflection of a single plasmon and the complete transmission peaks near the resonant frequencies, which is also a remarkable way to switch the scattering of a single plasmon. Our results can be utilized in the design of the next-generation quantum devices, such as single photon transistors, quantum switches and nanomirrors.

**Acknowledgments.** This work was supported by Key Project for Frontier Research on Quantum Information and Quantum Optics of Ministry of Education of D. P. R of Korea.

**References**

[1] A. Wallraff, D. I. Schuster, A. Blais, L. Frunzio, R.-S. Huang, J. Majer, S. Kumar, S. M. Girvin, and R. J. Schoelkopf, Nature (London) 431, 162 (2004); J. M. Raimond, M. Brune, and S. Haroche, Rev. Mod. Phys. 73, 565 (2001).

[2] P. Kok, W. J. Munro, K. Nemoto, T. C. Ralph, J. P. Dowling, and G. J. Milburn, Rev. Mod. Phys. 79, 135 (2007).

[3] D. E. Chang, A. S. Sørensen, E. A. Demler, and M. D. Lukin, "A single-photon transistor using nanoscale surface plasmons," Nat. Phys. **3**, 807 (2007).

[4] J.-T. Shen, S. Fan, Opt. Lett. 30 (2005) 2001; J.-T. Shen, S. Fan, Phys. Rev. Lett. 95, 213001(2005).

[5] J.-T. Shen, S. Fan, Phys. Rev. A 79, 023837(2009); J.-T. Shen, S. Fan, Phys. Rev. A 79, 023838(2009); M. Bradford, J-T. Shen, Phys. Rev. A 85, 043814 (2012).

[6] Z. R. Gong, H. Ian, Lan Zhou, and C. P. Sun, Phys. Rev. A 78, 053806 (2008); L. Zhou, H. Dong, Y. X. Liu, C. P. Sun, and F. Nori, Phys. Rev. A **78**, 063827 (2008); L. Zhou, Z. R. Gong, Y. X. Liu, C. P. Sun, F. Nori, Phys. Rev. Lett. 101, 100501 (2008).

[7] G. Zumofen, N. M. Mojarad, V. Sandoghdar, M. Agio, Phys. Rev. Lett. 101, 180404 (2008).

[8] G. Rostami, M. Shahabadi, A. Kusha, and A. Rostami, Applied Optics, 51, 5019 (2012).

[9] M. -T. Cheng, Y.-Q. Luo, P.-Z. Wang, and G.-X. Zhao, Appl. Phys. Lett.97, 191903 (2010).

[10] Yuecheng Shen and Jung-Tsung Shen, Phys. Rev. A 85, 013801 (2012).

[11] M.-T. Cheng, X.-S. Ma, M.-T. Ding, Y.-Q. Luo, and G.-X. Zhao, Phys. Rev. A 85, 053840 (2012).

[12] B. Dayan, A.S. Parkins, T. Aoki, E.P. Ostby, K.J. Vahala, H.J. Kimble, Science 319, 1062 (2008).

[13] Nam-Chol Kim, Jian-Bo Li, Shao-Ding Liu, Mu-Tian Cheng, and Zhong-Hua Hao, Chin. Phys. Lett. **27**, 034211 (2010).

[14] T. Aoki, B. Dayan, E. Wilcut, W.P. Bowen, A.S. Parkins, T.J. Kippenberg, K.J. Vahala, H.J. Kimble, Nature 443, 671 (2006).

[15] K. M. Birnbaum, A. Boca, R. Miller, A.D. Boozer, T.E. Northup, H.J. Kimble, Nature 436, 87(2005).

[16]T. S. Tsoi, C. K. Law, Phys. Rev. A 78, 063832 (2008); T. S. Tsoi, C. K. Law, Phys. Rev. A 80, 033823(2009).

[17]N.-C. Kim, J.-B. Li, Z.-J. Yang, Z.-H. Hao, and Q.-Q. Wang, Appl. Phys. Lett. 97, 061110 (2010).

[18]W. Chen, G.-Y. Chen, and Y.-N. Chen, Opt. Express 18, 10360 (2010).

[19]H. Wei and H.-X. Xu, Nanophotonics, 2, 155-169 (2013).

[20]D. E. Chang, A. S. Sørensen, P. R. Hemmer and M. D. Lukin, Phys. Rev. Lett. **97**, 053002 (2006).

[21]Y. Fedutik, V. V. Temnov, O. Schops, U. Woggon, M. V. Artemyev, Phys. Rev. Lett. **99**, 136802 (2007); H. M. Gong, L. Zhou, X. R. Su, S. Xiao, S. D. Liu, and Q. Q. Wang, Adv. Funct. Mater. **19**, 298 (2009).

[22] Zhong-Jian Yang, Nam-Chol Kim, Jian-Bo Li, Mu-Tian Cheng, Shao-Ding Liu, Zhong-Hua Hao, and Qu-Quan Wang, Optics Express **18**, 4006 (2010).

[23]A. V. Akimov, A. Mukherjee, C. L. Yu, D. E. Chang, A. S. Zibrov, P. R. Hemmer, H. Park, M. D. Lukin, Nature **450**, 402 (2007); H. Wei, D. Ratchford, X. Li, H. X. Xu, and C. K. Shih, Nano. Lett. **9**, 4168 (2009).